\documentclass[a4paper,fleqn,usenatbib]{mnras}

\usepackage{mathptmx}

\usepackage{ae,aecompl}

\usepackage{graphicx}	
\usepackage{amsmath}	
\usepackage{amssymb}	

\usepackage{color,epsfig} 
\usepackage[utf8]{inputenc} 

\usepackage[flushleft]{threeparttable}
\makeatletter
\newlength{\abovecaptionskip}%
\makeatother

\newcommand{\be}{\begin{equation}}
\newcommand{\ee}{\end{equation}}

\newcommand{\vect}[1]{\mathbf{#1}}

\def\msun{{\,{\rm M}_\odot}}
\def\rsun{{\,{\rm R}_\odot}}
\def\gcm3{\, \rm g \, cm^{-3}}

\def\G{\, \rm G}
\def\h{\, \rm h}
\def\d{\, \rm d}
\def\erg{\, \rm erg}

\def\rt{R_{\rm t}}

\def\rp{R_{\rm p}}

\def\mh{M_{\rm h}}

\def\mstar{M_{\star}}
\def\rstar{R_{\star}}

\def\msun{M_{\odot}}

\def\B{|\vect{B}|}
\def\tmag{t_{\rm mag}}
\def\tstr{t_{\rm str}}

\title[Magnetic field evolution in TDEs]{Magnetic field evolution in tidal disruption events}

\author[Clément Bonnerot, Daniel J. Price, Giuseppe Lodato and Elena M. Rossi]{Clément Bonnerot,$^{1}$\thanks{E-mail: bonnerot@strw.leidenuniv.nl}
Daniel J. Price,$^{2}$
Giuseppe Lodato,$^{3}$
and Elena M. Rossi$^{1}$
\\
$^{1}$Leiden Observatory, Leiden University, PO Box 9513, 2300 RA, Leiden, the Netherlands\\
$^{2}$Monash Centre for Astrophysics and School of Mathematical Sciences, Monash University, Clayton, Vic 3800, Australia\\
$^{3}$Dipartimento di Fisica, Università Degli Studi di Milano, Via Celoria, 16, Milano, 20133, Italy
}

\date{Accepted XXX. Received YYY; in original form ZZZ}

\pubyear{2017}

\begin{document}
\label{firstpage}
\pagerange{\pageref{firstpage}--\pageref{lastpage}}
\maketitle

\begin{abstract}
When a star gets tidally disrupted by a supermassive black hole, its magnetic field is expected to pervade its debris. In this paper, we study this process via smoothed particle magnetohydrodynamical simulations of the disruption and early debris evolution including the stellar magnetic field. As the gas stretches into a stream, we show that the magnetic field evolution is strongly dependent on its orientation with respect to the stretching direction. In particular, an alignment of the field lines with the direction of stretching induces an increase of the magnetic energy. For disruptions happening well within the tidal radius, the star compression causes the magnetic field strength to sharply increase by an order of magnitude at the time of pericentre passage. If the disruption is partial, we find evidence for a dynamo process occurring inside the surviving core due to the formation of vortices. This causes an amplification of the magnetic field strength by a factor of $\sim 10$. However, this value represents a lower limit since it increases with numerical resolution. For an initial field strength of 1 G, the magnetic field never becomes dynamically important. Instead, the disruption of a star with a strong 1 MG magnetic field produces a debris stream within which magnetic pressure becomes similar to gas pressure a few tens of hours after disruption. If the remnant of one or multiple partial disruptions is eventually fully disrupted, its magnetic field could be large enough to magnetically power the relativistic jet detected from Swift J1644+57. Magnetized streams could also be significantly thickened by magnetic pressure when it overcomes the confining effect of self-gravity.
\end{abstract}




\section{Introduction}
\label{introduction}

A tidal disruption event (TDE) happens when a star gets destroyed by the strong tidal forces of a supermassive black hole. Following the disruption, the stellar debris evolves into an extended stream of gas composed of a bound part that falls back towards the disruption site and an unbound part that escapes the black hole's gravity \citep{rees1988}. The central region of this stream can also contain a surviving self-gravitating core after the encounter if the star is only partially disrupted.

Stars commonly have a magnetic field that is expected to be transferred to the debris during a TDE. This magnetic field has several potentially interesting consequences on the debris subsequent evolution. Magnetic stresses within the stream can accelerate the circularization of its bound part into an accretion disc \citep{bonnerot2017-stream}. Alternatively, they can cause a fraction of the debris to pass beyond the event horizon of the black hole and be ballistically accreted \citep{svirski2017}. If field lines are oriented along the stream longitudinal direction, the associated magnetic tension can make the stream more resistant to hydrodynamical instabilities, predicted to otherwise affect the low-density streams produced by the disruption of giant stars \citep{mccourt2015,bonnerot2016-kh}. This magnetic effect could also prevent the stream fragmentation into self-gravitating clumps \citep{coughlin2015-variability}. Finally, while the stream is commonly thought to maintain a narrow profile set by hydrostatic equilibrium between gas pressure and self-gravity \citep{kochanek1994,coughlin2016-structure}, magnetic pressure could provide an additional outward force that can affect the stream structure, likely making it thicker than previously thought.

Among the few dozen TDE candidates detected so far, a small fraction shows evidence of a relativistic jet, the most famous example being Swift J1644+57 whose X-ray radiation is thought to be beamed along our line of sight \citep{bloom2011, burrows2011}. One mechanism to power a relativistic jet is the Blandford-Znajek mechanism \citep{blandford1977} which allows to extract rotational energy from the black hole. A necessary ingredient for this mechanism to operate is a large-scale magnetic field threading the black hole. The field lines then get twisted by the black hole rotation. As they unwind and expand, plasma gets ejected at high velocities along the direction of the black hole spin. However, in the case of Swift J1644+57, the stellar magnetic flux alone is far too small to launch a jet powerful enough to account for the measured X-ray luminosity \citep{tchekhovskoy2014}. Alternative origins have been proposed that involve an in-situ dynamo process creating regions of large magnetic flux within the disc \citep{krolik2012,piran2015-jet}, the interaction with a fossil disc whose magnetic field is collected by the stream in its fallback \citep{kelley2014} and the disruption of a strongly-magnetized star resulting from a recent binary merger \citep{mandel2015}. Another possibility is that TDE jets are powered radiatively \citep{sadowski2015,jiang2014,kara2016}, in which case a large magnetic flux is not required.

In this paper, we study the evolution of the stellar magnetic field as the star is tidally disrupted by a black hole by means of simulations using the smoothed particle magnetohydrodynamics (SPMHD) numerical method \citetext{see \citealt{price2012} for a review}, a generalization of the smoothed particle hydrodynamics (SPH) technique \citep{monaghan2005}. This approach is complementary to a recent study by \citet{guillochon2017-magnetic} which was carried out using a grid-based code and for different initial configurations. Depending on the orientation of the magnetic field with respect to the stream stretching direction, we find that the magnetic field distribution within the debris varies significantly. As expected from flux conservation, the stream magnetic field strength only slowly decreases if the field lines align with the direction of stretching resulting in a magnetic energy increase. For a partial disruption, we find evidence of a dynamo process occurring due to the formation of vortices at the interface between the surviving core and the recollapsing material. The magnetic field strength gets amplified within the core via this mechanism by about an order of magnitude. Instead, the star compression occurring for deep tidal disruptions lead to a sharp peak in the magnetic field strength at pericentre passage. Finally, the disruption of a strongly magnetized star results in a stream inside which magnetic pressure becomes comparable to gas pressure, providing an additional support against self-gravity.

The outline of the paper is as follows. In Section \ref{simulations}, we describe the numerical setup and method used to perform the simulations. The results are presented in Section \ref{results} which successively treats the influence of the magnetic field orientation, depth of the encounter and strength of the initial stellar magnetic field. The impact of the numerical resolution on these results is also evaluated. Finally, Section \ref{conclusion} contains a discussion of these results and our concluding remarks. In particular, we compare our results to that of \citet{guillochon2017-magnetic}.

\section{SPH simulations}
\label{simulations}

We simulate the interaction between a star of mass $\mstar=\msun$ and radius $\rstar=\rsun$ and a black hole of mass $ \mh=10^6 \msun$. For this choice of parameter, the tidal radius, within which the tidal force from the black hole exceeds the self-gravity force of the star, is $\rt = \rstar (\mh/\mstar)^{1/3}=100 \rsun$. The star is set on a parabolic orbit at a distance of $3 \rt$ from the black hole, where the tidal force represents only 4\% of the self-gravity force. Its pericentre distance $\rp$ is defined via the penetration factor $\beta\equiv\rt/\rp$, which we set to different values. We investigate $\beta=0.7$, for which the star is expected to be only partially disrupted with a surviving core continuing to orbit the black hole after the encounter \citep{guillochon2013}. Larger values of the penetration factor $\beta=1$ and $\beta=5$ are also considered that both correspond to a full disruption of the star. 

\begin{figure}
\epsfig{width=0.47\textwidth, file=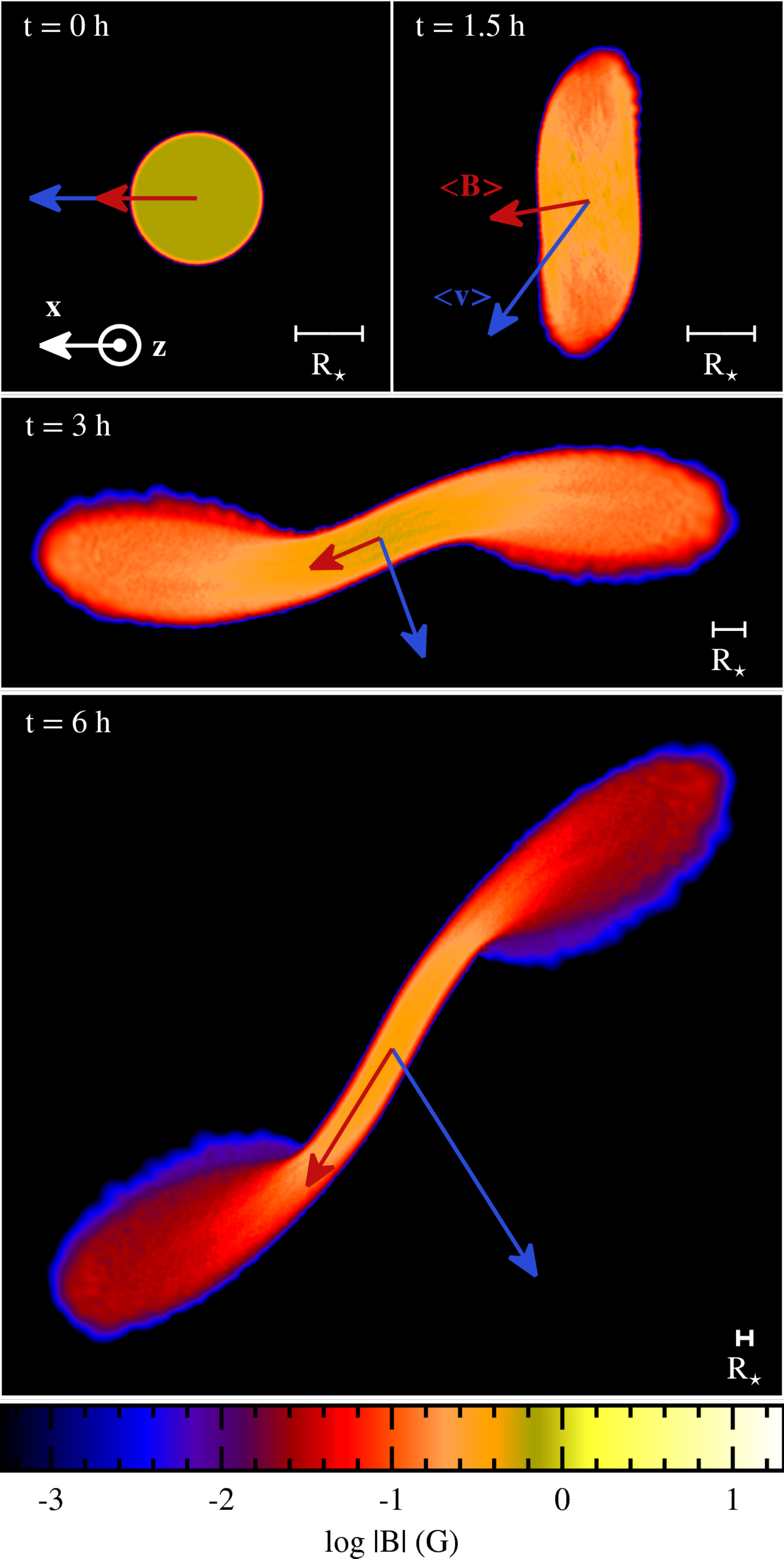}
\caption{Snapshots of the tidal disruption process showing the magnetic field strength of the gas for model F1B1G-x at different times $t=0$, 1.5, 3 and $6 \h$ in a reference frame that follows the centre of mass. The penetration factor is fixed to $\beta=1$. The blue arrows indicate the direction of the centre of mass velocity while the red ones represent the direction of the mean magnetic field. Their length does not have a physical meaning. The scale is different in each panel as indicated by the segment on the bottom right which represents the stellar radius. The white arrows on the upper left panel define the $\vect{x}$ and $\vect{z}$ directions. }
\label{fig1}
\end{figure}

The star is modelled as a polytropic sphere with $\gamma = 5/3$ containing one million SPH particles. A resolution study is presented in Section \ref{resolution} where different numbers of particles are considered. To achieve the desired density profile, the SPH particles are first positioned according to a close sphere packing and then differentially stretched along their radial direction. This structure is then evolved in isolation until its internal properties settle down. This technique has also been used by \citet{lodato2009}. In addition, an initial magnetic field is imposed to the star, which we choose to be uniform and linear for a clearer interpretation of the results, especially their dependence on varying magnetic field orientations. This choice is different from that of \citet{guillochon2017-magnetic} who consider a unique orientation of the field. Little is known about the strength of magnetic fields in stellar interiors. The magnetic field observed on stellar surfaces  have strengths varying between the solar value of $\sim 1 \G$ and  $\sim 10 \, \rm kG$ for rapidly rotating stars \citep{oksala2010}. In stellar interiors, evidence of magnetic fields with strengths of $\sim 1 \, \rm MG$ are found through asteroseismology measurements in red giants \citep{fuller2015}. We therefore adopt this range of values to model the stellar magnetic field in this paper. The dynamical importance of the magnetic field is measured by the plasma beta $\beta_{\rm M} \equiv P_{\rm gas}/P_{\rm mag}$, defined as the ratio of gas pressure to magnetic pressure. The latter is given as a function of magnetic field strength as $P_{\rm mag} \equiv \B^2/(8 \pi)$ cgs. The magnetic field strength is set to $|\vect{B}| = 1 \G$ in most of our models. For this choice, the initial plasma beta within the star is $\beta_{\rm M, ini} \approx 10^{16} \gg 1$, which implies that the magnetic field is not dynamically relevant. We investigate two different orientations of the field, pointing in the $\vect{x}$ and $\vect{z}$ directions illustrated on the upper left panel of Fig. \ref{fig1}. Note that the $\vect{x}$ direction is along the initial stellar orbit while the $\vect{z}$ direction is orthogonal to the orbital plane of the star. We also test the effect of increasing the magnetic field strength to $|\vect{B}| = 1 \, \rm MG$ and $|\vect{B}| = 2 \, \rm MG$, which corresponds to a strongly magnetized star. In this case, the initial plasma beta reaches $\beta_{\rm M, ini} \approx 10^{4}$. The magnetic fields considered are therefore never dynamically relevant initially. This justifies the method used to produce the initial condition where the magnetic field is added to the star after its evolution in isolation. Finally, we also perform a control simulation for which the star is not magnetized. The different models and the associated choice of parameters are summarized in table \ref{param}. 

The magnetic field is defined only on the SPH particles and not outside the star initially. A problem with this approach is that we therefore do not explicitly specify the boundary condition on the stellar magnetic field. In reality, the magnetic pressure outside the star far exceeds the ram pressure from the ambient medium, so the field lines should move freely with the star, which is what occurs in the simulation because the field is frozen to the fluid. In other words, the interior field evolves due to the deformation of the initial Lagrangian particle distribution, a mapping process that can be reproduced in post-processing since we find the dynamical influence to be unimportant. We thus preferred to leave the boundary condition free, and the similarity of our results to those shown by \citet{guillochon2017-magnetic} shows that this does not strongly affect the outcome.

The simulations are performed using the SPMHD code \textsc{phantom} \citep{price2010,lodato2010,price2017}. The self-gravity implementation makes use of a k-D tree algorithm \citep{gafton2011}. Direct summation is performed to handle short-range interactions according to an opening angle criterion with a critical value of 0.5. The magnetic field is evolved according to the constrained hyperbolic divergence cleaning algorithm developed by \citet{tricco2012} and \citet{tricco2016}. This algorithm imposes the condition $\nabla \cdot \vect{B} = 0$ in accordance with the non-existence of magnetic monopoles. This is achieved by imposing this divergence term to obey a damped propagation equation that efficiently reduces the divergence errors as they are transported with the fluid. With this technique, our divergence errors obey $h |\nabla \cdot \vect{B}|/|\vect{B}| < 0.1$ during the entire simulations where the left hand side is averaged on the SPH particles and $h$ denotes the smoothing length. In addition, the gas thermodynamical quantities are evolved according to an adiabatic equation of state. To accommodate for shocks, we make use of the standard artificial viscosity prescription in combination with the switch developed by \citet{cullen2010} to strongly reduce artificial viscosity away from shocks.

\begin{table}
\begin{threeparttable}
\centering
\caption{Name and parameters of the different models.}
\begin{tabular}{@{}llrrrrlrlr@{}}
\hline
Model \tnote{a} & Disruption & $\beta$ & Strength & Orientation\\
\hline
F1B0G & Full & 1 & $0 \G$ &  -- \\
F1B1G-x & Full & 1  & $1 \G$ &  $\vect{x}$ \\
F1B1G-z & Full & 1 &   $1 \G$ &  $\vect{z}$ \\
F1B1MG-x & Full & 1 &  $1 \, \rm MG$ &  $\vect{x}$ \\
F1B2MG-x & Full & 1 &  $2 \, \rm MG$ &  $\vect{x}$ \\
F5B1G-x & Full & 5 &   $1 \G$ &  $\vect{x}$ \\
P.7B1G-x & Partial & 0.7 & $1 \G$ &  $\vect{x}$ \\
\hline
\end{tabular}
\begin{tablenotes}
\item[a] The first letter ``F'' and ``P'' in the name of the models refer to full and partial tidal disruptions. The following two numbers indicates the value of the penetration factor $\beta$ and the magnetic field strength. Finally, the last letter refers to the magnetic field orientation.
\end{tablenotes}
\label{param}
\end{threeparttable}
\end{table}

\begin{figure*}
\epsfig{width=\textwidth, file=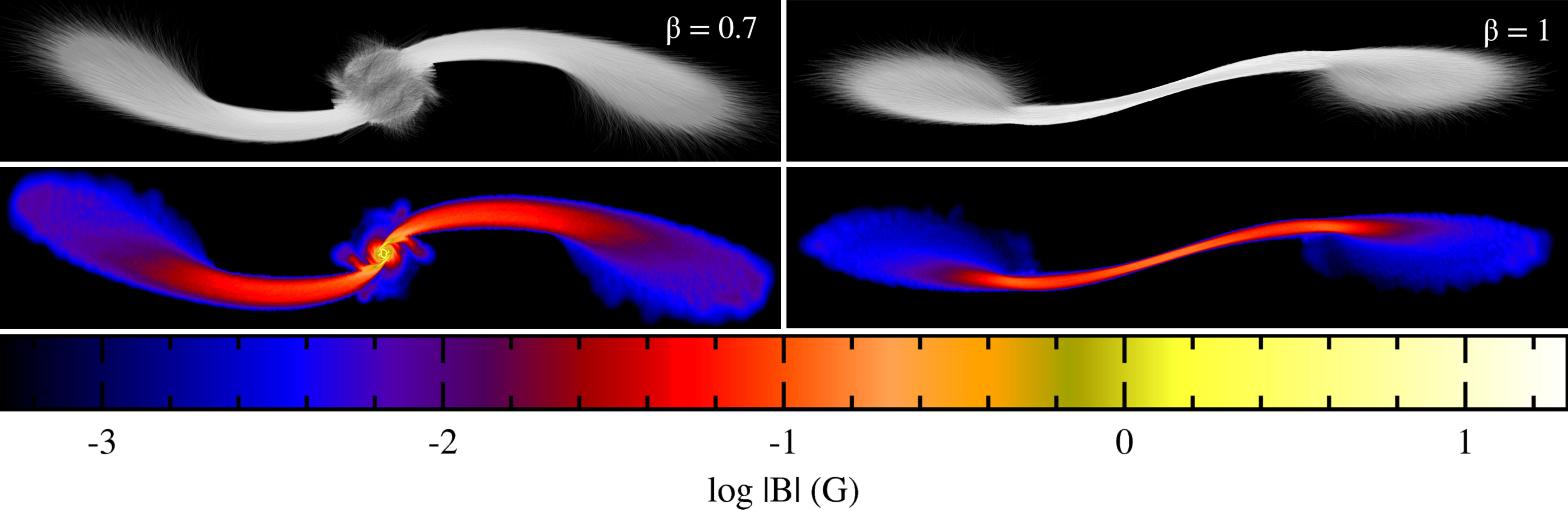}
\caption{Snapshots showing the magnetic field lines (upper panels) and strength (lower panels) at $t = 20 \h$ in the entire gas distribution for model P.7B1G-x (left panels) and F1B1G-x (right panels), for which the star is partially and fully disrupted respectively.}
\label{fig2}
\end{figure*}

\begin{figure}
\epsfig{width=0.47\textwidth, file=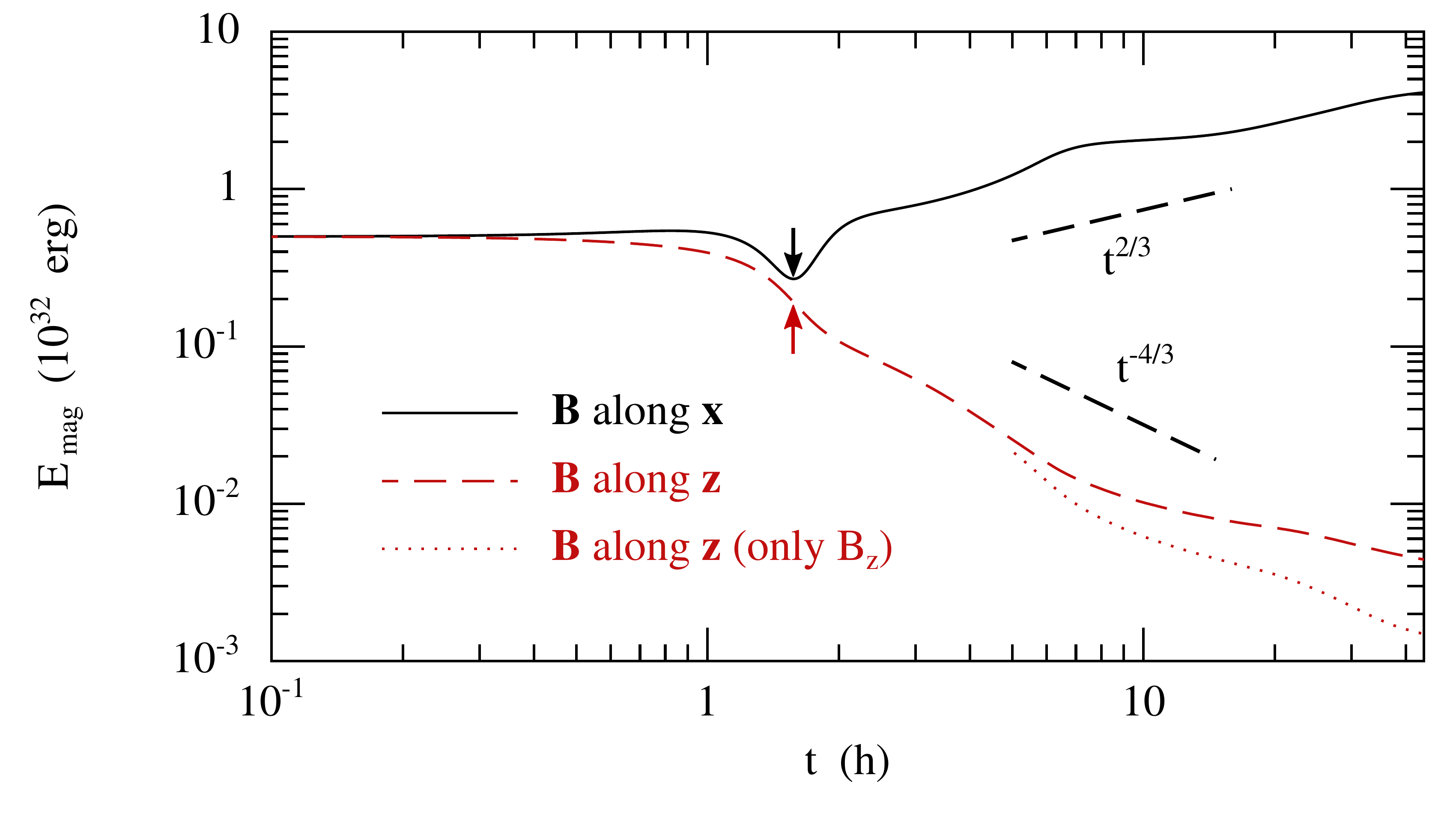}
\caption{Magnetic energy evolution for models F1B1G-x (black solid line) and F1B1G-z (red dashed line), for which the penetration factor is $\beta=1$. The dotted red line shows the magnetic energy computed from the $\bf{z}$ component of the field only for model F1B1G-z. The times of pericentre passage are indicated by the arrows on each curve. The two solid dashed segments indicate the scalings that the magnetic energy is expected to follow after the disruption.}
\label{fig3}
\end{figure}

\section{Results}
\label{results}

In this section, we present the results of the simulations.\footnote{Movies of the simulations presented in this paper are available at \url{http://home.strw.leidenuniv.nl/~bonnerot/research.html}.} The analysis is made by evaluating the evolution of the magnetic field strength of the debris and the total magnetic energy. The latter is given as a function of the field strength via $E_{\rm mag} \equiv \int P_{\rm mag} \d V \approx |\vect{B}|^2 V$ where $V$ represents the volume of the gas distribution. The initial magnetic energy is $E_{\rm mag} \approx 10^{32} \erg$ and $10^{44} \erg$ for $|\vect{B}| = 1 \rm G$ and $1 \rm MG$ respectively.

\subsection{Influence on the field orientation}
\label{orientation}

\begin{figure}
\epsfig{width=0.47\textwidth, file=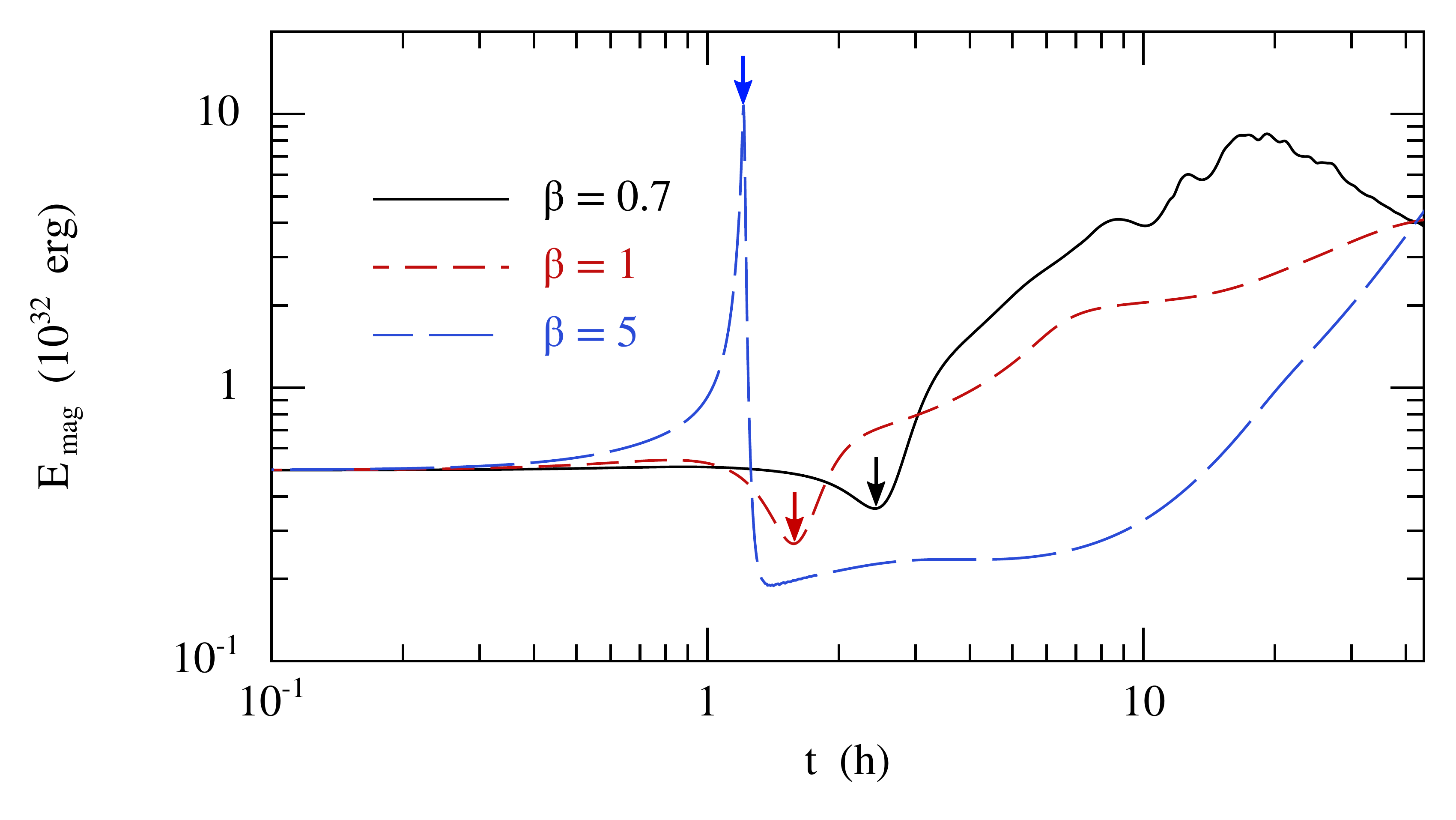}
\epsfig{width=0.47\textwidth, file=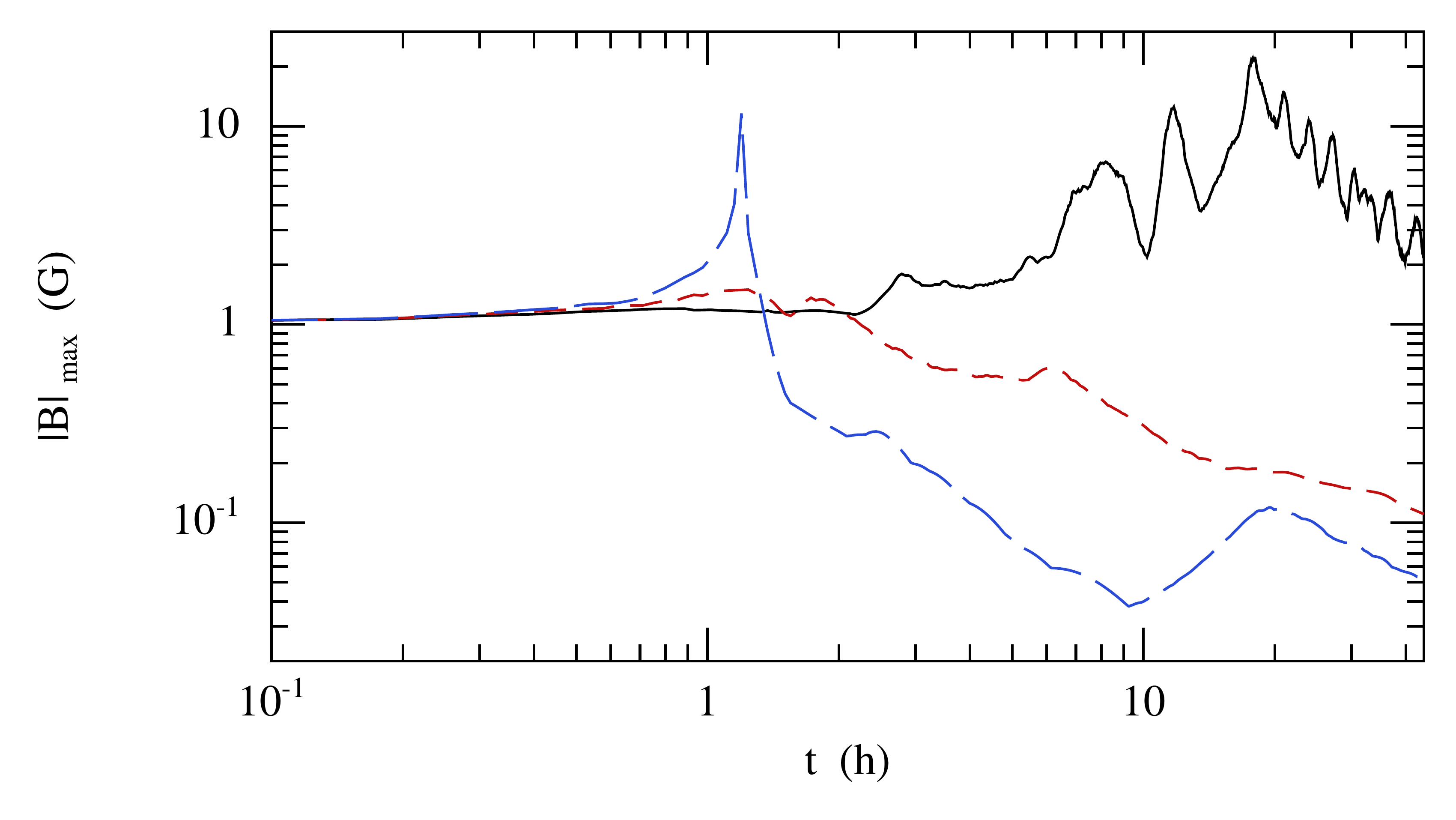}
\caption{Magnetic energy (upper panel) and maximal magnetic field strength (lower panel) evolution for models P.7B1G-x (black solid line), F1B1G-x (red dashed line) and F5B1G-x (long-dashed blue line). The times of pericentre passage are indicated by the arrows on each curve.}
\label{fig4}
\end{figure}

\begin{figure*}
\epsfig{width=\textwidth, file=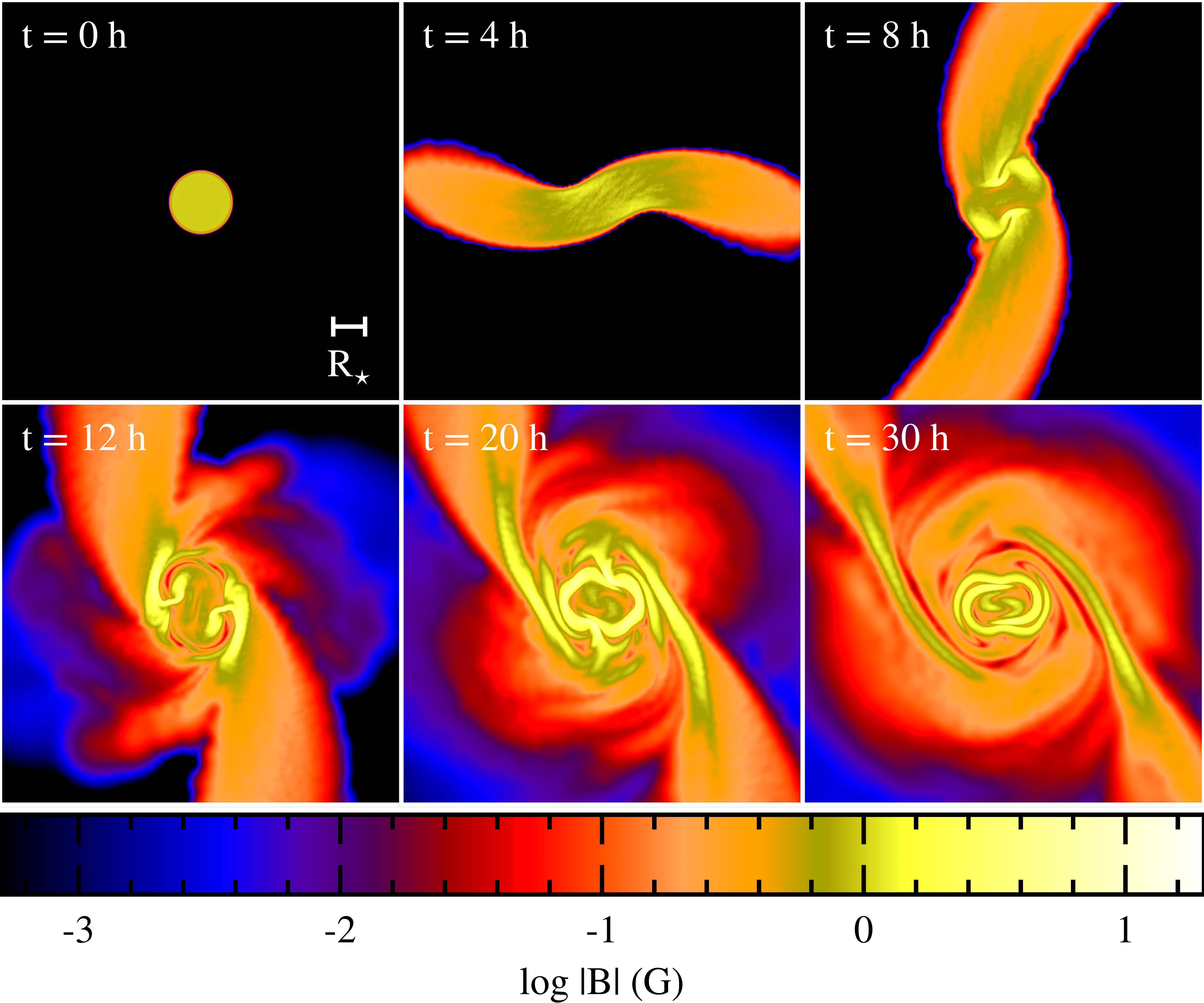}
\caption{Snapshots showing the magnetic field strength at different times $t = 0$, 4, 8, 12, 20 and $30 \h$ for model P.7B1G-x, corresponding to a partial disruption of the star. The reference frame follows the centre of mass of the gas distribution. The star reaches pericentre at $t \approx 2.5 \h$. The magnetic field gets amplified within the surviving core to values up to $|\vect{B}| \approx 10 \G$}
\label{fig5}
\end{figure*}

\begin{figure}
\epsfig{width=0.47\textwidth, file=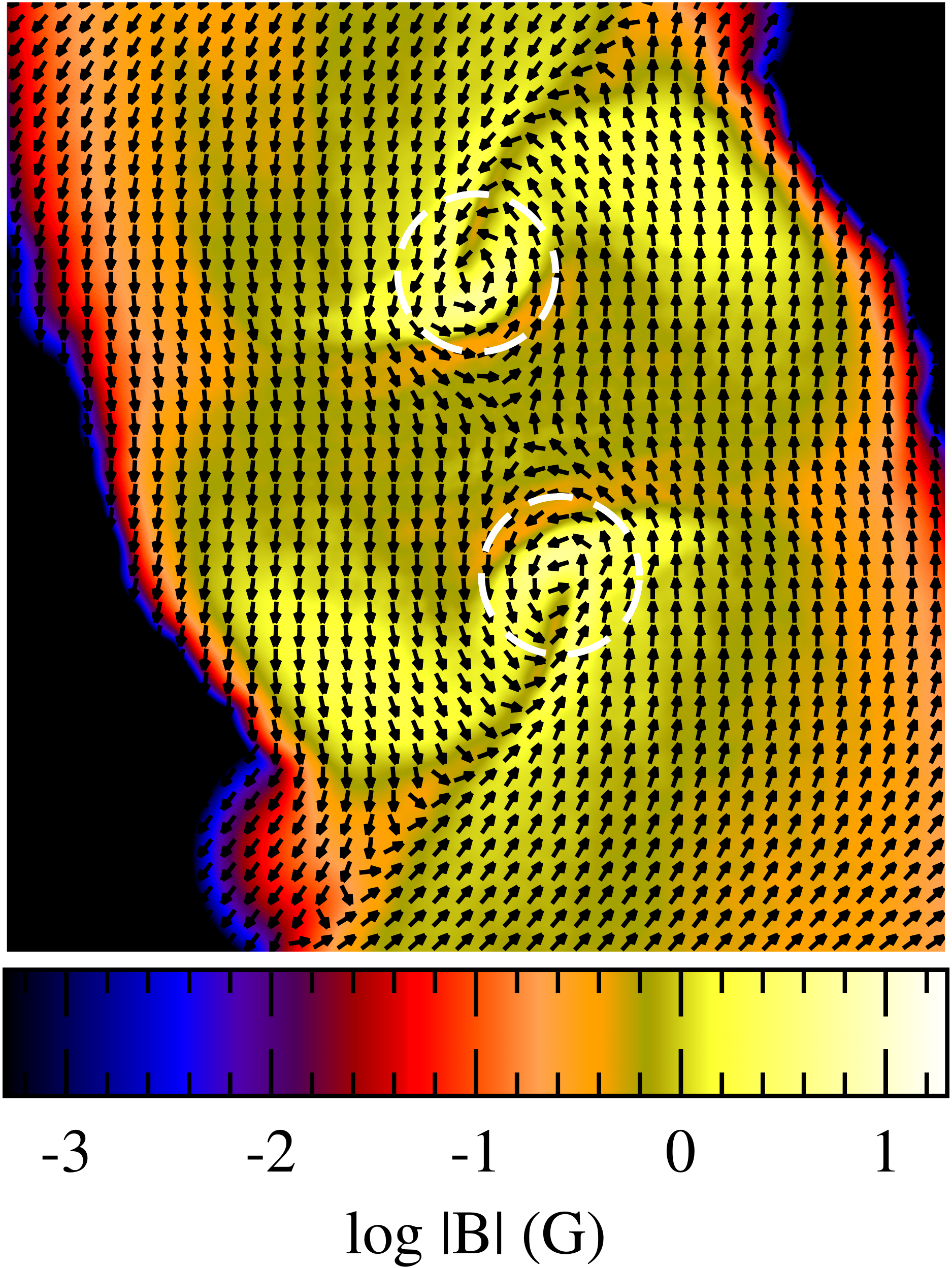}
\caption{Close-up on the centre of mass of the gas distribution showing the magnetic field strength at $t = 8 \h$ for model P.7B1G-x. The arrows denote the velocity field, which features two zones of rotational motion corresponding to vortices. The location of these vortices correspond to the zones of largest magnetic field strength.}
\label{fig6}
\end{figure}

First, we evaluate the impact of the stellar magnetic field orientation on its distribution within the debris, focusing on full disruptions with a fixed penetration factor $\beta=1$ and magnetic field strength $|\vect{B}| = 1 \rm G$. For this purpose, we compare models F1B1G-x and F1B1G-z, for which the stellar magnetic field is linear and oriented in the $\vect{x}$ and $\vect{z}$ directions respectively. Recall that the $\vect{x}$ direction is aligned with the initial trajectory of the star while the $\vect{z}$ direction is orthogonal to the orbital plane.

The hydrodynamics is indistinguishable between the two models since the magnetic field is dynamically irrelevant owing to the large value of the plasma beta $\beta_{\rm M}\simeq 10^{16} \gg 1$. The gas evolution is presented in Fig. \ref{fig1} which shows snapshots of the tidal disruption process following the centre of mass of the star at different times $t=0$, 1.5, 3 and 6 h. The colours represent the magnetic field strength for model F1B1G-x while the arrows indicate the direction of the centre of mass velocity (blue arrow) and the direction of the mean magnetic field (red arrow). Initially, the orientation of the magnetic field is imposed by the initial conditions. At $t=1.5 \h$, the star reaches pericentre where it gets stretched by a factor of $\sim 2$ due to the velocity difference between material on each side of the trajectory, the gas closer to the black hole moving faster than the matter further away. This elongation takes place in the direction perpendicular to the magnetic field for model F1B1G-x. The stellar debris then evolves into a stream that keeps stretching at later times. For model F1B1G-x, the magnetic field gets re-oriented in the direction of stretching as the stream continues to expand as can be noticed from the red arrows on the two lowermost panels of Fig. \ref{fig1}. This is because, in absence of magnetic diffusion, each magnetic field line must pass through the same fluid elements at all times. Therefore, as the gas distribution gets stretched, so do the field lines causing the magnetic field to re-orient in the stretching direction. For this model (F1B1G-x), the magnetic field lines orientation and strength within the debris are shown in the right panel of Fig. \ref{fig2} at $t = 20 \h$. It illustrates the alignment of the field lines with the direction of stretching and the magnetic strength mild decrease to an average of $|\vect{B}| \approx 0.1 \rm G$

For model F1B1G-z, the direction of the magnetic field is unaffected by the gas evolution and remains in the initial $\vect{z}$ direction. Also in this case, the field lines are frozen in the flow and follow the stream elongation. However, since the star's stretching occurs in the orbital plane, this is not accompanied by a re-orientation of the field lines. The magnetic field therefore remains orthogonal to the direction of stretching at all times.

The evolution of the magnetic energy is shown in Fig. \ref{fig3} for models F1B1G-x (solid black line) and F1B1G-z (dashed red line) with the time of pericentre passage indicated by an arrow on each curve. Using the gas and magnetic field evolution described above, each trend can be understood from magnetic flux conservation that imposes $|\vect{B}| \propto 1/S$, where $S$ is the surface orthogonal to the field direction. The magnetic energy therefore scales as $E_{\rm mag} \approx |\vect{B}|^2 \, V \propto V/S^2$. For model F1B1G-x, the magnetic energy drops slightly at the moment of disruption. This is due to the elongation experienced by the star in the direction orthogonal to the magnetic field, seen in the upper right panel of Fig. \ref{fig1}. Afterwards, the energy increases until the end of our simulation. This increase is caused by the stretching of the stellar debris in the direction parallel to the magnetic field, also visible on the two lowermost panels of Fig. \ref{fig1}. The rate of increase can be understood as follows from magnetic flux conservation. The surface through which the field lines pass is orthogonal to the stretching direction and scales as $S_{\perp}\propto H^2$ while the volume of the gas distribution evolves like $V \propto H^2 l$. $H$ and $l$ represent the width and length of a fluid element belonging to the stream respectively. This implies that the magnetic energy evolves as $E_{\rm mag} \propto V/S^2_{\perp} \propto l/H^2$. Using the scalings $H \propto t^{1/3}$ and $l \propto t^{4/3}$ derived by \citet{coughlin2016-structure} during this phase of evolution, the magnetic energy scales as $E_{\rm mag} \propto t^{2/3}$. This scaling is indicated by the upper black dashed segment in Fig. \ref{fig3} and provides an accurate description of the magnetic energy evolution for model F1B1G-x. Note that, even if the magnetic energy increases, the magnetic field strength decreases since $\B \propto 1/H^2 \propto t^{-2/3}$ as illustrated in the bottom panel of Fig. \ref{fig4} (red dashed line).\\

For model F1B1G-z, the evolution is significantly different since the magnetic energy decreases during the whole simulation. This is because, as explained above, the magnetic field remains orthogonal to the stretching direction in this case. The surface parallel to the stream stretching scales as $S_{\parallel} \propto l H$. As a result, magnetic flux conservation imposes $E_{\rm mag} \propto V/S^2_{\parallel} \propto 1/l$. Since $l \propto t^{4/3}$, the magnetic energy evolves as $E_{\rm mag} \propto t^{-4/3}$ . As can be seen from Fig. \ref{fig3} by comparing the dashed red line to the lower black dashed segment, the magnetic energy follows this scaling closely for model F1B1G-z. At $t \gtrsim 5 \h$, the magnetic energy can however be seen to decrease slightly slower than the scaling. This is due to small components of the magnetic field along the orbital plane originating from the shearing experienced by the debris during the tidal disruption process. These additional components of the magnetic field tend to increase the total magnetic energy, making the decrease slower than expected. This interpretation is demonstrated by computing the magnetic energy including only the $\bf{z}$ component of the magnetic field. As can be seen from the red dotted line in Fig. \ref{fig3}, this partial magnetic energy follows the expected scaling. At late times, the magnetic energy is small enough to be affected by the presence of low-density regions where the magnetic field is overestimated due to divergence errors, with $h |\nabla \cdot \vect{B}|/|\vect{B}| \gtrsim 0.1$. This artificially causes the magnetic energy to reach a plateau at $t \gtrsim 20$ h. The SPH particles leading to this unphysical behaviour have densities three orders of magnitude lower than the mean and represent only $\sim 1\%$ of the whole distribution. They have been removed to compute the magnetic energy shown in Fig. \ref{fig3} for model F1B1G-z.

It can also be noticed from Fig. \ref{fig3} that the magnetic energy evolution for both models F1B1G-x and F1B1G-z slightly differs from the above scalings at $t \approx 7 \h$ where it experiences a small oscillation, also seen in the density evolution. This density oscillation has already been identified in the simulations performed by \citet{coughlin2016-pancakes} and was also found to happen around 5.5 h after pericentre passage for the set of parameters considered here (see their figure 8). It is triggered by a compression of the debris along the orbital plane due to the differential motion of the front and back of the star at the moment of disruption. The density variation is accompanied by a modification of the stream profile causing $H$ to increase slightly slower than the previous scaling and $l$ slightly faster. As a result, the magnetic energy $E_{\rm mag} \propto l/H^2$ for model F1B1G-x increases faster, creating a bump. Similarly, the magnetic energy $E_{\rm mag} \propto 1/l$ for model F1B1G-z decreases faster, producing a hollow.

\begin{figure}
\epsfig{width=0.47\textwidth, file=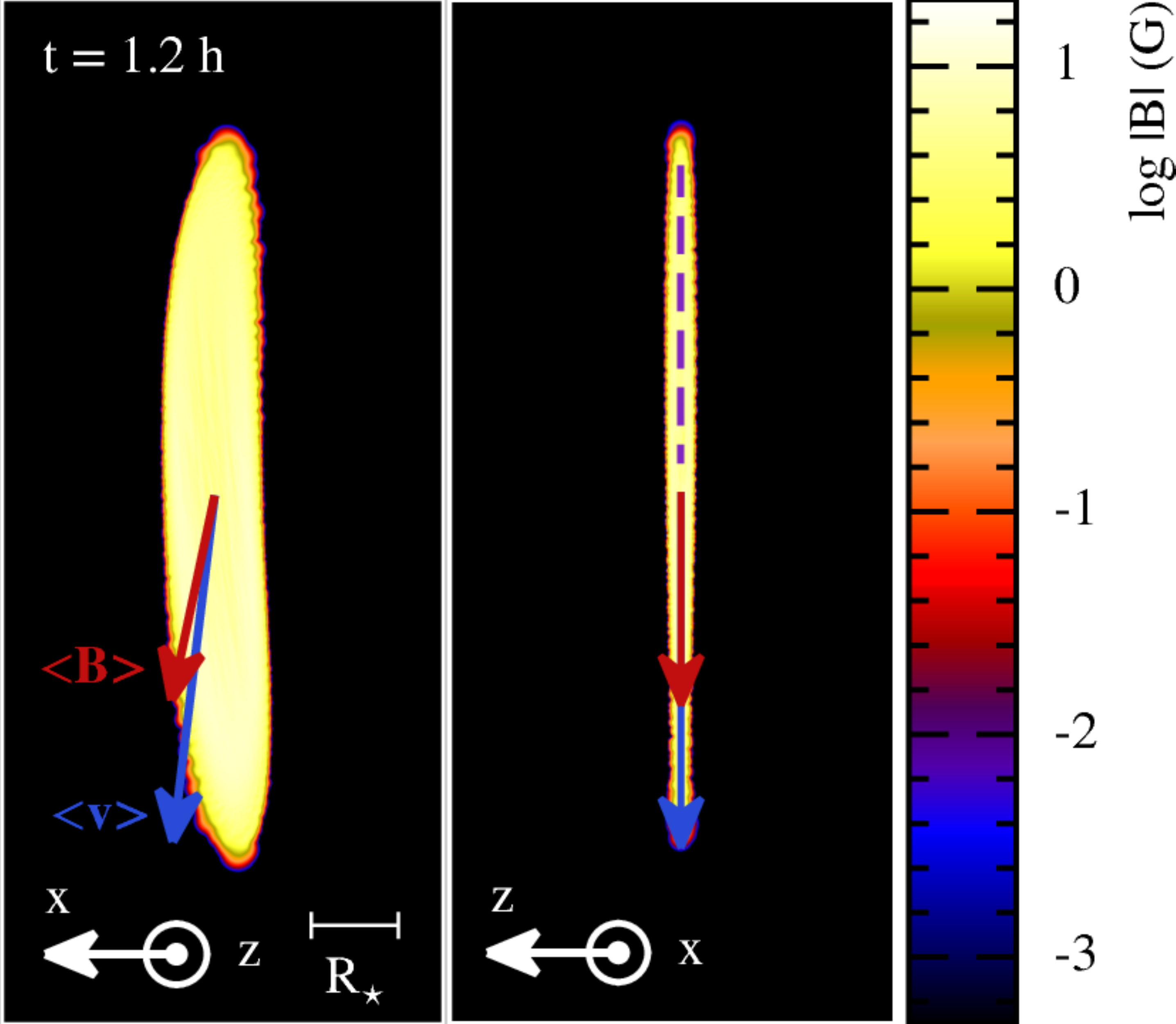}
\caption{Magnetic field strength of the gas for model F5B1G-x at the time of pericentre passage $t = 1.2 \h$ along a line of sight orthogonal (left panel) and parallel (right panel) to the orbital plane of the star. The blue arrows indicate the direction of the centre of mass velocity while the red ones represent the direction of the mean magnetic field. On the right panel, the vertical dashed purple segment indicates the orbital plane of the star.}
\label{fig7}
\end{figure}

\subsection{Dependence on the penetration factor}

We investigate the effect of the penetration factor on the magnetic field evolution by comparing models P.7B1G-x and F5B1G-x with model F1B1G-x, already discussed in Section \ref{orientation}. For model P.7B1G-x, the penetration factor is fixed to $\beta = 0.7$, for which the disruption is expected to be partial. It is increased to $\beta = 1$ and 5 for models F1B1G-x and F5B1G-x, both leading to full disruptions. These three models adopt the same initial magnetic field strength of $|\vect{B}|=1 \G$ and a common orientation along the $\vect{x}$ direction.

We look at model P.7B1G-x first. Fig. \ref{fig5} shows the gas evolution and its magnetic field strength for this model in a frame of reference following the centre of mass. The star reaches pericentre at $t \approx 2.5$ h, after which it gets stretched to form an elongated structure. This initial phase of evolution is similar to that of model F1B1G-x, shown on the three uppermost panels of Fig. \ref{fig1}. However, the subsequent evolution differs due to the lower value of the penetration factor $\beta = 0.7$. Starting from $t \approx 8$ h, matter starts to collapse towards the centre of mass to form a self-gravitating core. The interaction of this re-collapsing gas with the rotating core leads to the formation of two vortices close to the core surface. It is important to notice that these vortices have a purely hydrodynamical origin since the magnetic field is never dynamically relevant in our simulation. These features can be seen by looking at Fig. \ref{fig6} which shows a close-up on the surviving core at $t=8 \h$ where the velocity field computed in the reference frame of the centre of mass is indicated as black arrows. Clearly, this velocity field exhibits two zones of rotational motion, highlighted by the dashed white circles. Inside these vortices, the magnetic field gets amplified to reach strengths up to $|\vect{B}| \approx 10 \G$. Note that even after this amplification, the magnetic field remains dynamically irrelevant in the core with a plasma beta of $\beta_{\rm M} \approx 10^{13} \gg 1$. Later in time, the core keeps rotating causing the formation of a more complex magnetic structure as can be seen in Fig. \ref{fig5} for $t \geq 12$ h. We also notice that the rotational motion associated to the vortices progressively disappears until the only gas motion identifiable in the core is that due to its rigid rotation.

This evolution is fundamentally different from that of model F1B1G-x described in Section \ref{orientation}, for which the disruption was total. A comparison can be made by looking at Fig. \ref{fig2} that shows the magnetic field lines and strength within the whole gas distribution at $t=20 \h$ for models P0.7B1G-x (left panels) and F1B1G-x (right panels). The magnetic field strength is comparable for the two models away from the centre of mass of the stellar debris, where it is always $|\vect{B}|\lesssim 0.1 \G$. The field lines are also similar, directed along the stream longitudinal direction. Instead, the magnetic field structure differs significantly near the centre of mass. At this location, model P.7B1G-x features a complex magnetic configuration due to the formation of a self-gravitating core.  The magnetic field strength gets amplified and the field lines become tangled. For model F1B1G-x, there is instead no of magnetic field amplification and the field lines are directed along the stretching direction everywhere through the debris.

Fig. \ref{fig7} shows the gas distribution and its magnetic field strength for model F5B1G-x when the star reaches pericentre, at $t = 1.2 \h$. As in Fig. \ref{fig1}, the left panel adopts a line of sight orthogonal to the orbital plane. On the right panel, the orbital plane is indicated by the dashed purple segment along the gas distribution and the line of sight is parallel to it. This allows to see the gas elements above and below the orbital plane. The red arrow denotes the direction of the mean magnetic field while the blue arrow shows the centre of mass velocity. The star gets elongated along its orbital plane by a factor of $\sim 8$ as it passes at pericentre. This elongation is analogous to that seen for model F1B1G-x. However, it is more pronounced due to the larger penetration factor $\beta = 5$ that causes the star to pass closer to the black hole where tidal forces are stronger. This larger elongation for model F5B1G-x has a consequence on the magnetic field evolution. As can be seen from Fig \ref{fig7}, the magnetic field gets re-oriented in the direction of elongation by the time of pericentre passage. This re-orientation of the field lines has the same origin as for model F1B1G-x, discussed in Section \ref{orientation}. However, it occurs earlier due to the larger elongation factor. At pericentre passage, the mean magnetic field is still close to its initial orientation for model F1B1G-x (upper right panel of Fig. \ref{fig1}) but it is already re-oriented along the direction of stretching for model F5B1G-x. The right panel of Fig. \ref{fig7} shows that the star is additionally compressed by factor of $\sim 3$ in the direction orthogonal to its orbital plane. This strong vertical collapse is expected for deep-penetrating encounter, for which the matter passes well within the tidal radius \citep{carter1983,stone2013}.

A more quantitative analysis can be done using Fig. \ref{fig4} that shows the magnetic energy (upper panel) and maximal magnetic field strength (lower panel) for models P.7B1G-x (solid black line), F1B1G-x (dashed red line) and F5B1G-x (long-dashed blue line). The times of pericentre passage, different for each model, are indicated by the arrows on each curve. For model P.7B1G-x, the magnetic energy increases the fastest shortly after the disruption due to the dynamo process at play in the surviving core. It scales as $E_{\rm mag} \propto t^p$ where $p \approx 1.4$ compared to $E_{\rm mag} \propto t^{2/3}$ for model F1B1G-x. The maximal magnetic field strength also increases to reach $\B_{\rm max} \approx 20\G$ at $t \approx 18 \h$. At $t\gtrsim 20 \h$, the magnetic energy starts decreasing. However, this late stage of evolution appears to be strongly resolution dependent and will be discussed later in Section \ref{resolution}. For model F1B1G-x, the magnetic field strength continuously decreases down to $\B \approx 0.1\G$ while the magnetic energy increases. For model F5B1G-x, the evolution is similar except for a large peak at the time of pericentre passage where the maximal magnetic field strength reaches $\B_{\rm max} \approx 10\G$. This is due to the strong compression experienced by the star in the direction perpendicular to its orbital plane (right panel of Fig. \ref{fig7}). Since the magnetic field is orthogonal to the direction of compression, flux conservation imposes an associated increase of the magnetic field strength which explains the peak seen in Fig. \ref{fig4} for model F5B1G-x (long-dashed blue line). At later times, the evolution is similar to that of model F1B1G-x.

\subsection{Impact of the field strength}

We now focus on the impact of the field strength on the debris evolution by analysing model F1B1MG-x, for which the stellar magnetic field strength is increased to $\B= 1 \, \rm MG$. This is six orders of magnitude larger than for model F1B1G-x discussed in Section \ref{orientation}, where the strength was $\B= 1 \, \rm G$. However, the initial field remains oriented in the $\vect{x}$ direction and the penetration factor is fixed to $\beta=1$.

\begin{figure}
\epsfig{width=0.47\textwidth, file=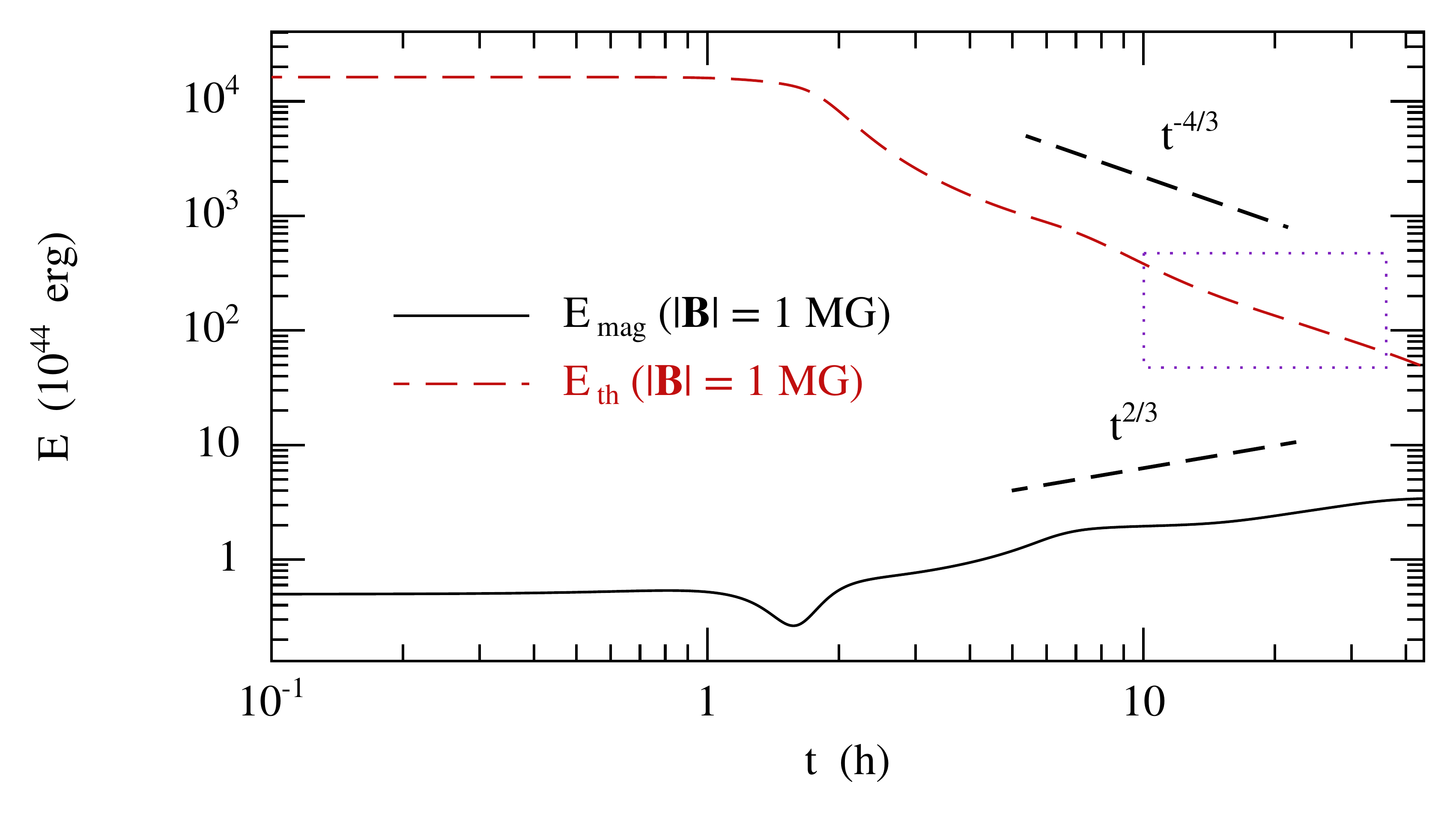}
\epsfig{width=0.47\textwidth, file=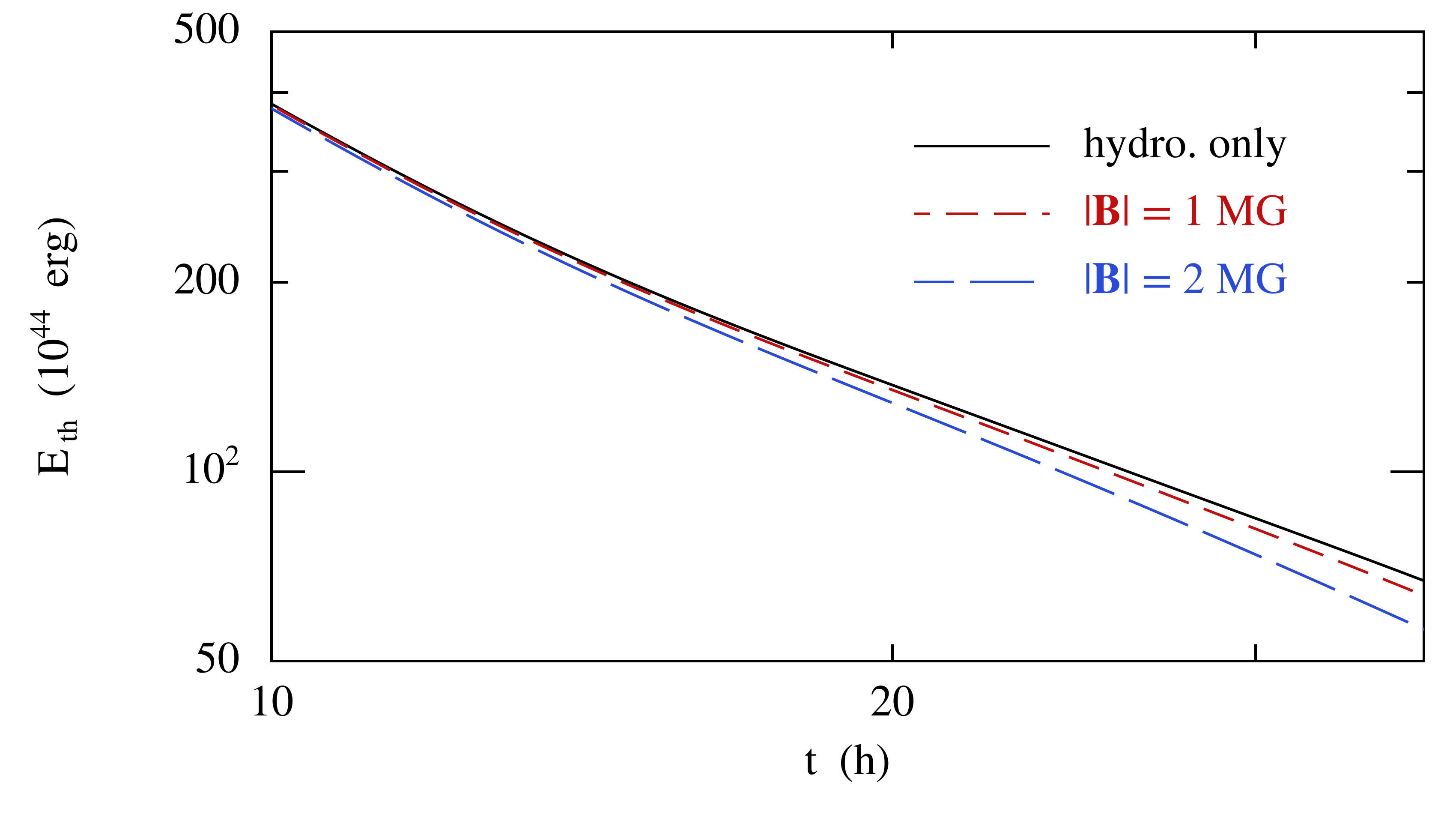}
\caption{Evolution of the magnetic (black solid line) and thermal (red dashed line) energies for model F1B1MG-x (upper panel). The dashed black segments indicate the scaling that these energies are expected to follow after the disruption. The area delimited by the dotted purple rectangle is zoomed-in on the lower panel, which shows the late time thermal energy evolution for models F1B0G (solid black line), F1B1MG-x (dashed red line) and F1B2MG-x (long-dashed blue line).}
\label{fig8}
\end{figure}

Fig. \ref{fig8} (upper panel) shows the evolution of the magnetic (black solid line) and thermal (red dashed line) energies for model F1B1MG-x. As expected, the magnetic energy evolution is identical to that of model F1B1G-x with an energy increase that follows $E_{\rm mag} \propto t^{2/3}$. It is only shifted upwards by twelve orders of magnitude owing to the larger initial magnetic field strength. On the other hand, the thermal energy decreases after disruption due to the expansion of the stream. This energy is given by $E_{\rm th} \equiv (3/2) \int P_{\rm gas} \d V   \approx P_{\rm gas} V$. Since the evolution is adiabatic, the gas pressure scales as $P_{\rm gas} \propto \rho^{5/3} \propto V^{-5/3}$ where $\rho \propto 1/V$ represents the gas density. As a result, $E_{\rm th} \propto V^{-2/3} \propto t^{-4/3}$ using $V=H^2 l$ and the temporal dependence of $H$ and $l$ derived by \citet{coughlin2016-structure}. This slope is indicated by the upper dashed black segment in Fig. \ref{fig8} (upper panel) that is followed closely by the thermal energy. After the disruption, the magnetic energy therefore approaches the thermal energy until, at $t \approx 20$ h, they only differ by an order of magnitude. By that time, the plasma beta $\beta_{\rm M} \approx E_{\rm th}/E_{\rm mag}$ has decreased by three orders of magnitude, from $\beta_{\rm M, ini} \approx 10^4$ initially to $\beta_{\rm M} \approx 10$. This suggests that magnetic pressure is starting to have an dynamical impact on the stream structure. The area indicated by the dotted purple rectangle is zoomed-in on the lower panel of Fig. \ref{fig8}, which shows the late time evolution of the thermal energy for model F1B1MG-x (red dashed line) compared to the control model F1B0G (solid black line) with hydrodynamics only and model F1B2MG-x (blue long-dashed line) for which the star has a larger initial magnetic field strength of $\B = 2 \,\rm MG$. The thermal energy is reduced for increasing magnetic field strengths compared to the non-magnetized case. We interpret this variation as the effect of magnetic pressure that provides an additional support to thermal pressure against self-gravity to ensure hydrostatic equilibrium. To test this interpretation, we compare the variation $\Delta E_{\rm th}$ in thermal energy to the magnetic energy. For both models F1B1MG-x and F1B2MG-x, the ratio of these two quantities is found to be $\Delta E_{\rm th}/E_{\rm mag} = 1.02 \approx 1$ at $t = 20 \h$, which confirms that the decrease in thermal energy compared to the non-magnetized case is due to the presence of magnetic pressure. For  model F1B1G-x where the magnetic field strength is of only $\B = 1 \,\rm G$, we find that the thermal energy is identical to model F1B0G, meaning that the magnetic field does not affect the stream structure at any time.

This late-time impact of the magnetic pressure only results from the evolution of magnetic and thermal energies, which increases and decreases respectively during to the stream stretching. This effect is therefore general to every tidal disruption of magnetized stars as long as the magnetic field has an initial component in the direction of stretching. The latter condition is necessary to ensure an increase of the magnetic energy as demonstrated in Section \ref{orientation}.  In this situation, the magnetic pressure is expected to become significant at a finite time $\tmag$ after disruption. Since the plasma beta satisfies $\beta_{\rm M} \equiv E_{\rm th}/E_{\rm mag} \propto t^{-2}$, this characteristic timescale is given by
\be 
\tmag = \tstr \, \beta^{1/2}_{\rm M, ini},
\label{tmag}
\ee
where $\tstr$ denotes the stretching timescale, after which the stream has expanded by a significant amount. As physically expected, the magnetic pressure becomes significant earlier for more magnetized stars since $\tmag$ increases with $\beta_{\rm M, ini}$. The stretching timescale can be obtained from $\tstr=\rstar/\Delta v$ where $\Delta v$ denotes the velocity difference within the stellar debris imparted by tidal forces at the time of pericentre passage. For $\beta \approx 1$, $\Delta v \approx (G \mstar/ \rstar)^{1/2}$ and the stretching timescale is simply the stellar dynamical time, $\tstr = 0.4 \h$  for a solar-type star. This is consistent with the time delay found in our simulations between the disruption of the star and a significant stretching of the debris. Injecting this expression into equation \eqref{tmag} leads to
\be
t_{\rm mag} = 44 \h \left(  \frac{\beta_{\rm M, ini}}{ 10^4} \right)^{1/2}  \left(  \frac{\mstar}{\msun} \right)^{-1/2} \left(  \frac{\rstar}{\rsun} \right)^{3/2},
\label{tmag-value}
\ee
consistent with the time at which the magnetic pressure becomes comparable to the gas pressure.

\subsection{Resolution study}
\label{resolution}

\begin{figure}
\epsfig{width=0.47\textwidth, file=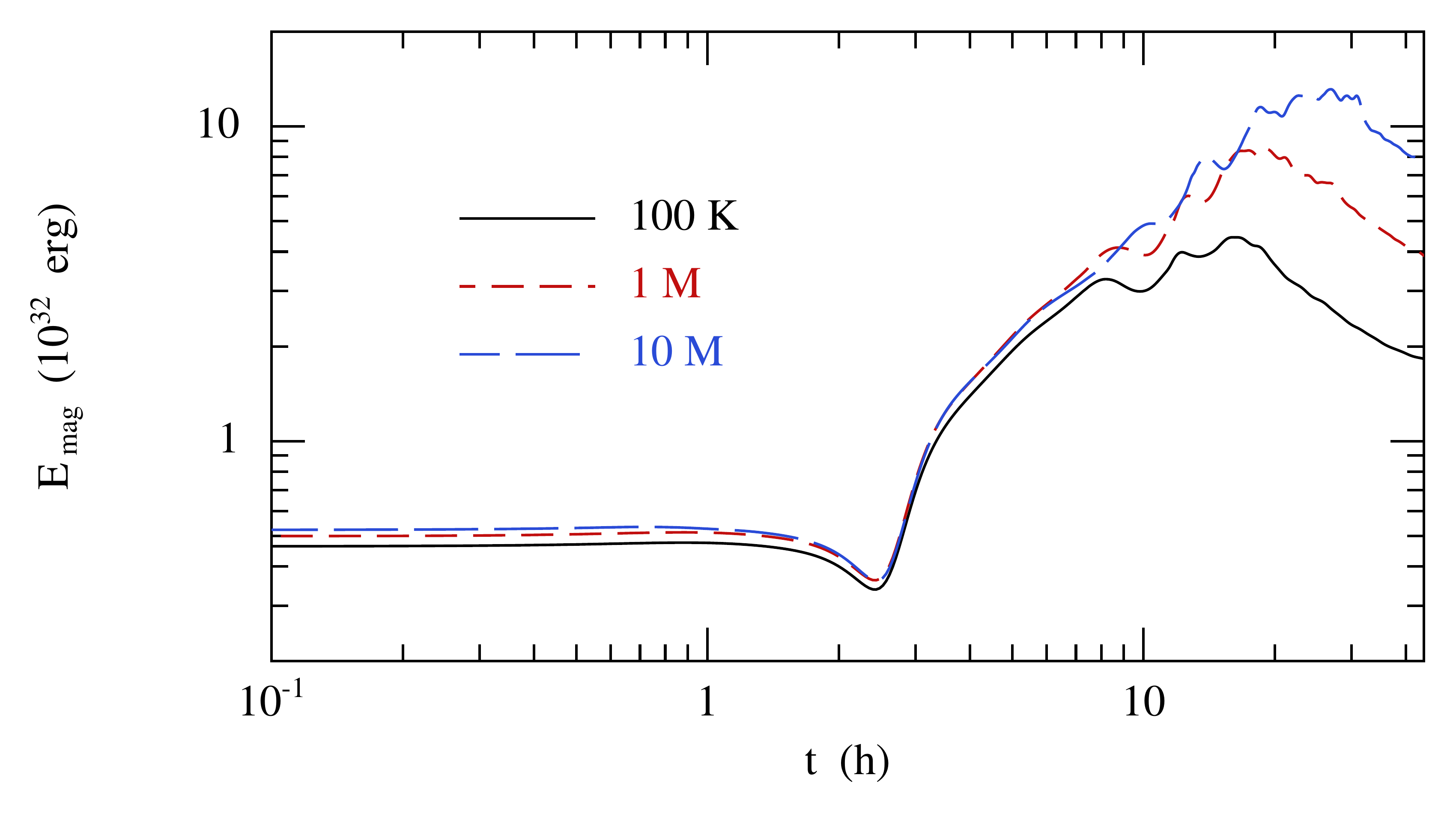}
\caption{Magnetic energy evolution for model P0.7B1G-x using $10^5$ (black solid line), $10^6$ (red dashed line) and $10^7$ (blue long-dashed line) SPH particles for the simulation.}
\label{fig9}
\end{figure}

We now evaluate the effect of numerical resolution on the results of our simulations. This is done by focusing on model P0.7B1G-x such that both the magnetic field evolution imposed by stream stretching and the dynamo process at play in the surviving core can be analysed. The magnetic energy evolution is shown in Fig. \ref{fig9} for model P0.7B1G-x adopting three different numbers of SPH particles: $10^5$ (black solid line), $10^6$ (dashed red line) and $10^7$ (long-dashed blue line). Small differences in the initial magnetic energy can be noticed between different resolutions. They are only due to slight variations in the volume of the initial particle distribution within the star. The magnetic energy evolution close to pericentre passage and shortly after is identical for the three resolutions. For $t \gtrsim 4$ h, the initial magnetic field growth is the same for the two largest resolutions but the magnetic energy starts to differ for the lowest resolution. Up to this time, our simulations have therefore already reached convergence for $10^6$ particles, the number used for the results presented in this paper. When $t \gtrsim 10$ h, the magnetic energy significantly diverges for the three resolutions. Magnetic field amplification is sustained for a longer time at higher resolutions which results in a larger peak value for the magnetic energy. Between the two lower (larger) resolutions, the peak in magnetic energy is delayed by $\sim 3 \h$ ($\sim 7.6 \h$) and larger by 89\% (54\%). We connect this longer magnetic field amplification observed at higher resolution to the fact that the vortices within which the dynamo process operates are longer-lived. Given the dependence on resolution, we interpret this effect as being due to numerical dissipation. At higher resolution, numerical dissipation is reduced and the vortices disappear later in time. Therefore, we conclude that the magnetic field amplification seen in our simulations must be understood as a lower limit. A physical upper limit will be estimated in Section \ref{conclusion}.

\section{Discussion and conclusion}
\label{conclusion}

The evolution of the magnetic field of the star during its tidal disruption has not received significant attention despite its potentially fundamental importance. In this paper, we perform magnetohydrodynamical simulations of the tidal disruption process considering the stellar magnetic field. We find that the initial magnetic field orientation significantly affects the post-disruption magnetic energy evolution because it determines the inclination of the magnetic field with respect to the stream stretching direction. As expected from flux conservation, the magnetic field strength of the debris decreases slowly when the field lines are aligned with the stretching direction resulting in an increase of the magnetic energy. Instead, the magnetic energy decreases when the magnetic field is perpendicular to the direction of stretching. We also find that varying the depth of the encounter leads to qualitative differences in the magnetic field evolution. For a deeply penetrating encounter, the magnetic field strength undergoes a sharp increase close to pericentre passage caused by a strong compression of the star. Instead, a partial disruption leads to the formation of a surviving core inside which vortices form. We find clear evidence of a dynamo process at play within these vortices, which induces an increase of the magnetic field strength by about an order of magnitude. For the disruption of strongly magnetized stars, we show that magnetic pressure provides an additional support against self-gravity in the stream transverse direction after a few tens of hours for an initial magnetic field strength $\B \geq 1 \, \rm MG$. This action of magnetic pressure is also to be expected for less magnetized stars, but on a longer timescale (equation \eqref{tmag-value}).

In our simulations, we find that magnetic pressure provides an additional although marginal support to gas pressure against self-gravity. Since the magnetic energy can increase with time, a possibility is that, in highly magnetized stars, magnetic pressure may exceed self-gravity at later times. As a result, the width of the stream would no longer be confined by self-gravity but would become thicker with a transverse profile entirely determined by magnetic pressure. Several other mechanisms have so far been proposed to counteract the effect of self-gravity which include thermal energy injection at the moment of disruption for large penetration factors $\beta \gtrsim 3$ and hydrogen recombination, which occurs about a week after the disruption \citep{kasen2010}. The thickening effect of magnetic pressure on the stream could also affect its subsequent circularization. In particular, disc formation might not be delayed by Lense-Thirring precession, thought to prevent an early self-crossing shock for narrow streams revolving around spinning black holes \citep{guillochon2015}. In addition, circularization can be accelerated since magnetic stresses are able to strengthen self-crossing shocks during the disc formation process \citep{bonnerot2017-stream}.

Hydrodynamical instabilities can significantly affect the low-density stream of debris expected from tidal disruptions involving giant stars or massive black holes. However, these instabilities are prevented by magnetic tension if the stream has a longitudinal magnetic field component of strength $|\vect{B_{\parallel}}| \approx 1-10 \G$ \citep{bonnerot2016-kh}. In our simulations, the magnetic field lines naturally align with the stream longitudinal direction as long as the initial magnetic field has a component along the stretching direction. If stars host strongly-magnetized cores with $|\vect{B}| \gg 1 \G $, even a small fraction of their magnetic field would therefore be enough to prevent these instabilities from developing. This possibility is favoured by recent evidence for large magnetic fields in the cores of red giants with strengths $|\vect{B}| \gtrsim 10^5 \G$, obtained from asteroseismology \citep{fuller2015}.

In the case of a partial disruption, the amplification factor found in our simulations is only a lower limit since higher resolution simulations produce longer-lived vortices due to decreased numerical dissipation, suggesting that at even higher resolution the vortices could result in long-lived hydrodynamic turbulence. An upper limit for the amplification factor can be obtained from equipartition between the core rotational energy and its magnetic energy. This upper limit can be reached only if a sustained dynamo develops which remains unaffected until equipartition. In practice, the dynamo process is likely to be stopped earlier by various physical processes such as internal dissipation within the surviving core. A detailed study of the internal structure of the surviving core is therefore required to determine the exact amount of magnetic field amplification, which is beyond the scope of this paper but could be carried out by means of a stellar evolution code. The core rotational energy is $E_{\rm rot} \approx \mstar \Delta v^2/2 \approx 10^{48} \erg$, sixteen orders of magnitude larger than the magnetic energy for $|\vect{B}| = 1 \G$. Equipartition would therefore induce an amplification of the magnetic field strength of the core up to $|\vect{B}| \approx 10^8 \G$. This implies that strongly-magnetized stellar cores may naturally result from partial tidal disruptions. If, as our simulations find, a sustained dynamo does not develop, a large magnetic field amplification could still be reached if the star experiences a series of several partial disruptions during which its magnetic field is mildly amplified. Starting from a stellar magnetic field of $|\vect{B}| = 1 \G$, the magnetic field strength reached in the stellar core after $N_{\rm p}$ pericentre passages is $|\vect{B}| = f^{N_{\rm p}}_{\rm amp} \G$ which results in
\be
N_{\rm p} = \frac{\log|\vect{B}|}{\log f_{\rm amp}}.
\ee
Using the value $f_{\rm amp} \approx 10$ found in our simulations for the amplification factor, a star therefore needs to experience $N_{\rm p} = 8$ pericentre passages for its core to reach a magnetic field strength of $|\vect{B}| = 10^8 \G$. If a star is disrupted after a strong magnetic field amplification, the magnetic field flux brought by the stellar debris could be sufficient to power the relativistic jets detected from a fraction of TDEs. For Swift J1644+57, the required magnetic field strength has been estimated to $|\vect{B}| \approx 10^8 \G$ \citep{tchekhovskoy2014} that could be achieved either after a single pericentre passage if a sustained dynamo takes place within the core or after $\sim 8$ encounters using the lower limit on the amplification factor given by our simulations. However, since the field lines align with the stream longitudinal direction, the newly-formed disc could lack the poloïdal magnetic field component required for jet launching. Theoretically, partial disruptions are expected to represent between $\sim 20 \%$ and the large majority of all TDEs depending on the regime of angular momentum relaxation into the loss cone \citep{stone2016-rates}. Such events are also proposed to account for the low value of the total radiated energy obtained from numerous observations of TDEs \citep[e.g.][]{chornock2014}. Specifically, the scenario of a full tidal disruption following one or multiple partial disruptions is favoured if the star slowly diffuses into the loss cone through small changes of its angular momentum \citep[section 4]{strubbe2011}. However, the remnant may also avoid a subsequent total disruption if it is scattered off its orbit by a two-body encounter \citep{alexander2001}. Hydrodynamical effects are also likely to affect this picture. After a partial disruption, the surviving core can get unbound from the black hole due to asymmetric mass loss \citep{manukian2013,gafton2015}. On the other hand, heating of the surviving core at pericentre is done at the expanse of its orbital energy and could make it expand to be more easily disrupted at the next passage close to the black hole \citep{cheng2014}. In addition, the stellar core trajectory might be affected by its interaction with the mass lost from previous encounters present close to the black hole. Detailed hydrodynamical simulations of successive partial disruptions are necessary to determine the dominant effect.

For deep-penetrating encounters, the magnetic field strength is found to peak due to compression at pericentre. The associated increase of magnetic pressure could result in an additional support against compression that is likely to impact the subsequent bounce, computed by considering gas pressure only \citep{stone2013}.

Several investigations of magnetic field amplification during neutron star and white dwarf mergers have been carried out. In this context, both SPH \citep{price2006} and moving-mesh \citep{zhu2015} simulations tend to result in magnetic field amplifications larger by orders of magnitude than in grid-code simulations \citep{kiuchi2014}. In \citet{price2006}, the fast growth was an artefact of a boundary condition effect from using the Euler potentials. The method used by \citet{zhu2015} does not include divergence cleaning which likely explains the large magnetic field amplification seen in their simulations. In the present study, we find an amplification of the magnetic field consistent with the recent grid code simulations performed by \citet{guillochon2017-magnetic} thanks to the divergence cleaning method used to reduce divergence errors \citep{tricco2012,tricco2016}. We found in some of our early calculations that turning off the divergence cleaning could produce spurious dynamo amplification on timescales similar to those found by \citet{zhu2015}.

We provided a study of the stellar magnetic field evolution during the tidal disruption of a star and early debris evolution. In the future, we aim at investigating the longer-term effect of the magnetic field on the debris, especially its impact on the stream internal structure and dynamical influence during the circularization process.

\section*{Acknowledgments}

We thank the anonymous referee for useful comments which improved the paper.
CB is grateful to the Monash Centre for Astrophysics for hosting him during the completion of this work. He also thanks Mark Avara, Roseanne Cheng, James Guillochon, Yuri Levin, David Liptai, Christopher Matzner, Rebecca Nealon, Josiah Schwab and Nicholas Stone for interesting discussions. CB and EMR acknowledge the help from NOVA. DJP is grateful for funding via Future Fellowship FT130100034 from the Australian Research Council. Finally, we used the SPH visualization tool SPLASH \citep{price2007} for producing all the figures of this paper.




\bibliographystyle{mnras} 
\bibliography{biblio}




\bsp	
\label{lastpage}
\end{document}